\newcommand{\CLASSINPUTtoptextmargin}{0.85in}
\newcommand{\CLASSINPUTbottomtextmargin}{0.95in}
\documentclass[conference]{IEEEtran}
\IEEEoverridecommandlockouts
\usepackage{standalone}%

\usepackage{cite}
\usepackage{amsmath,amssymb,amsfonts}
\usepackage{mathtools}
\usepackage{bm}
\usepackage{algorithmic}
\usepackage{graphicx}
\usepackage{textcomp}
\usepackage{subcaption}
\usepackage{xcolor}
\def\BibTeX{{\rm B\kern-.05em{\sc i\kern-.025em b}\kern-.08em
    T\kern-.1667em\lower.7ex\hbox{E}\kern-.125emX}}

\usepackage{trfsigns}
\usepackage{placeins}
\graphicspath{{inkscape/}} %
\usepackage{tabularx}         
\usepackage{array,multirow}  
\usepackage{hhline}                    
\usepackage[shortcuts,acronym]{glossaries}
\usepackage{blkarray, bigstrut}
\newcommand\myvec{\boldsymbol}%
\newcommand\mymat[1]{\boldsymbol{\mathbf{#1}}}

\newcommand\myvar[1]{\mathsf{\MakeUppercase{#1}}}
\newcommand\mylabel[1]{\mathrm{#1}}

\newcommand{\Mod}[1]{(\mathrm{mod} #1)}
\usepackage{pgfplots}
\pgfplotsset{compat=newest} %
\usepackage{tikz}
\usetikzlibrary{calc}
\pgfkeys{
    /includestandaloneresized/.is family, /includestandaloneresized,
    default/.style = {
        width = \textwidth,
        ratio = 0.61803398875},
    width/.estore in = \itgWidth,
    ratio/.estore in = \itgRatio
}
\newlength\figurewidth
\newlength\figureheight

\usepackage{tikz-3dplot}
\usetikzlibrary{shapes,arrows,perspective,intersections,positioning,calc,automata,decorations.pathreplacing,decorations.markings, chains,fit,backgrounds, spy, matrix,decorations.pathreplacing}
\makeatletter
\def\ps@IEEEtitlepagestyle{%
	\def\@oddfoot{\mycopyrightnotice}%
	\def\@evenfoot{}%
}
\def\mycopyrightnotice{%
	{\footnotesize
		This work has been submitted to the IEEE for possible publication. Copyright may be transferred without notice, after which this version may no longer be accessible.\hfill}%
	\gdef\mycopyrightnotice{}%
}
\begin{document}
\setlength\abovecaptionskip{0.45\baselineskip}
\setlength{\textfloatsep}{0.35\baselineskip}
\def\@IEEEfigurecaptionsepspace{\vskip\abovecaptionskip\relax}%
\def\@IEEEtablecaptionsepspace{\vskip\abovecaptionskip\relax}%
\title{Low-Resolution Horizontal and Vertical Layered Mutual Information Maximizing LDPC Decoding}

\author{
\IEEEauthorblockN{Philipp Mohr, Gerhard Bauch}
\IEEEauthorblockA{\textit{Hamburg University of Technology} \\
\textit{Institute of Communications}\\
21073 Hamburg, Germany \\
Email: \{philipp.mohr, bauch\}@tuhh.de}
}

\maketitle

\newacronym{app}{APP}{a-posteriori probability}
\newacronym{bp}{BP}{belief propagation}
\newacronym{llr}{LLR}{log-likelihood ratio}
\newacronym{lut}{LUT}{lookup table}
\newacronym{luts}{LUTs}{lookup tables}
\newacronym{ib}{IB}{information bottleneck}
\newacronym{ldpc}{LDPC}{low-density parity-check}
\newacronym{qc}{QC}{quasi-cyclic}
\newacronym{omsq}{OMSQ}{quantized offset-min-sum}
\newacronym[longplural={variable nodes}]{vn}{VN}{variable node}
\newacronym[longplural={check nodes}]{cn}{CN}{check node}

\begin{abstract}
%auto-ignore
We investigate iterative low-resolution message-passing algorithms for quasi-cyclic LDPC codes with horizontal and vertical layered schedules. Coarse quantization and layered scheduling are highly relevant for hardware implementations to reduce the bit width of messages and the number of decoding iterations. As a novelty, this paper compares the two scheduling variants in combination with mutual information maximizing compression operations in variable and check nodes. We evaluate the complexity and error rate performance for various configurations. Dedicated hardware architectures for regular quasi-cyclic LDPC decoders are derived on a conceptual level. The hardware-resource estimates confirm that most of the complexity lies within the routing network operations. Our simulations reveal similar error rate performance for both layered schedules but a slightly lower average iteration count for the horizontal decoder.

\end{abstract}

%auto-ignore
\section{Introduction}
Mutual information maximizing low-density parity-check (LDPC) decoders have recently been shown to outperform conventional algorithms when using coarse resolutions for the messages exchanged between variable and check nodes~\cite{lewandowsky_information-optimum_2018, he_mutual_2019, monsees_finite-alphabet_2022}. 
Yet, only a few works address decoding with a layered schedule~\cite{mohr_coarsely_2021, kang_generalized_2022, wang_reconstruction-computation-quantization_2022}, which is of great practical relevance as it can halve the number of required decoding iterations compared to the flooding scheme~\cite{hocevar_reduced_2004, zhang_shuffled_2005}. In particular we are not aware of results on comparing horizontal and vertical scheduling in combination with mutual information maximizing decoders. 

\begin{figure}[h]
     \centering
	% This file was created by tikzplotlib v0.9.8.
%    \resizebox{\figurewidth}{\figureheight}{
%	\tikzset{every picture/.style={line width=0.6pt}}
	\def\unit{1.55cm}  % here you set units used in image
	\begin{tikzpicture} [scale=1.0, x=1*\unit,y=1*\unit, node distance=1.5*\unit and 1.5*\unit,
		wire/.style={black},
		mygray/.style={color=gray!70, text=gray!70},
		cn style/.style={rectangle, fill=black,minimum width=0.2*\unit, minimum height=0.2*\unit},
		vn style/.style={circle, minimum width=0.25*\unit}]   
		\pgfmathsetmacro\nrows{2};		
		\pgfmathsetmacro\ncols{3};
		\pgfmathsetmacro\Z{40};
		\pgfmathsetmacro\deltaZ{1/\Z};
		\def \basemat {{{0,0},{0,13},{0,27}}}
		% create matrix
		\foreach \r in {1,...,\nrows}{
			\foreach \c in {1,...,\ncols}{
%				\coordinate(A) at (vn1\c.west|-cn1\r.east);
				\coordinate(ul\r\c) at (\c-1,-\r+1); % upper left
				\coordinate(lr\r\c) at (\c,-\r); % lower right
				\coordinate(ur\r\c) at (\c,-\r+1); % upper right
				\coordinate(ll\r\c) at (\c-1,-\r); % lower left
				\node(submat\r\c)[draw, rectangle, fit=(ul\r\c)(lr\r\c), inner sep=0]{};
				\pgfmathtruncatemacro{\rm}{\r-1};
				\pgfmathtruncatemacro{\cm}{\c-1};
				\pgfmathtruncatemacro{\baseshift}{\basemat[\cm][\rm]};
				\foreach \k in {1,...,\Z}{
					\pgfmathsetmacro\sr{\k};
					\pgfmathsetmacro\sc{Mod(\k-1+\baseshift,\Z)+1};
					\draw[black, fill=black] ($(ul\r \c)+(\sc*\deltaZ-\deltaZ,-\sr*\deltaZ)$) rectangle ($(ul\r \c)+(\sc*\deltaZ,-\sr*\deltaZ+\deltaZ)$);
				}
			}
		}
		
		% create variable nodes
		\foreach \c in {1,...,\ncols}{
				\node(vn1\c) at (ul1\c)[draw, vn style, xshift=0.2*\unit, yshift=0.3*\unit]{};
				\node(vnZ\c) at (ur1\c)[draw, vn style, xshift=-0.2*\unit, yshift=0.3*\unit]{};
				\node(vdots\c) at ($(vn1\c)!0.5!(vnZ\c)$)[mygray]{\scriptsize$\dots$};
				% wires
				\foreach\r in {1,Z}{
				\draw[wire]([xshift=0.05*\unit]submat1\c.north-|vn\r\c.south)--(vn\r\c.south);
				\draw[wire]([xshift=-0.05*\unit]submat1\c.north-|vn\r\c.south)--(vn\r\c.south);
				}
			
				\node(col layer \c)[draw, orange, rectangle, rounded corners,fit=(vn1\c)(vnZ\c), inner sep=2pt]{};
				\node(col label \c) at (col layer \c)[orange, minimum width=0.8*\unit,align=center, xshift=0.0*\unit,yshift=0.42*\unit, inner sep=0pt]{vertical \\layer \c};
		}

		% create check nodes
		\foreach \r in {1,...,\nrows}{
			\node(cn\r 1) at (ur\r\ncols)[draw, cn style, xshift=0.3*\unit, yshift=-0.2*\unit]{};
			\node(cn\r Z) at (lr\r\ncols)[draw, cn style, xshift=0.3*\unit, yshift=0.2*\unit]{};
			\node(cdots\r) at ($(cn\r 1)!0.5!(cn\r Z)$)[rotate=90]{\scriptsize$\dots$};
			% wires
			\foreach\c in {1,Z}{
				\draw[wire]([yshift=0.08*\unit]submat\r\ncols.east|-cn\r \c.west)--(cn\r \c.west);
				\draw[wire]([yshift=0.0*\unit]submat\r\ncols.east|-cn\r \c.west)--(cn\r \c.west);
				\draw[wire]([yshift=-0.08*\unit]submat\r\ncols.east|-cn\r \c.west)--(cn\r \c.west);
			}
			
			\node(row layer \r)[draw, teal, rectangle, rounded corners,fit=(cn\r 1)(cn\r Z), inner sep=2pt]{};
			\node(row label \r) at (row layer \r)[teal, minimum width=0.8*\unit,align=center, yshift=-0.0*\unit,xshift=0.7*\unit]{horizontal\\layer \r};
%			\node(label};
%			\node[text width=1cm,align=center] {0,1,2\\3,4,5};
		}	
		% vertical orange lines
		\foreach \c in {2,...,\ncols}{
			\pgfmathtruncatemacro{\cm}{\c-1};
			\draw[thick,orange](ur1\cm|-col label 1)--(lr\nrows\cm);
		}
		% horizontal green lines
		\foreach \r in {2,...,\nrows}{
			\pgfmathtruncatemacro{\rm}{\r-1};
			\draw[thick,teal](ll\rm1)--(lr\rm\ncols-|row label 1);
		}
		\node(hmat) at ($(ul11)!0.5!(ll\nrows1)$)[xshift=-0.4*\unit]{$\mymat{H}=$};
		
		\draw[latex-latex]([yshift=-0.1cm]ll\nrows1)--([yshift=-0.1cm]lr\nrows\ncols);
		\node at (lr\nrows\ncols)[yshift=-0.1cm, xshift=0.1*\unit, inner sep = 0pt]{$N$};
		
		\draw[latex-latex]([xshift=-0.1cm]ul11)--([xshift=-0.1cm]ll\nrows1);
		\node at (ul11)[yshift=-0.1cm, xshift=-0.2*\unit, inner sep = 0pt]{$M$};
		
		\draw[latex-latex]([xshift=0.1cm,yshift=-0.2cm]ul11)--([xshift=0.1cm,yshift=0.2cm]ll11);
		\node at ($(ll11)!0.5!(ul11)$)[xshift=0.15*\unit, inner sep = 0pt]{$Z$};
	\end{tikzpicture}
	%}
    \caption{Illustration of horizontal and vertical layers.}
    \label{fig:pcmqcldpcverticalhorizontal}
\end{figure}
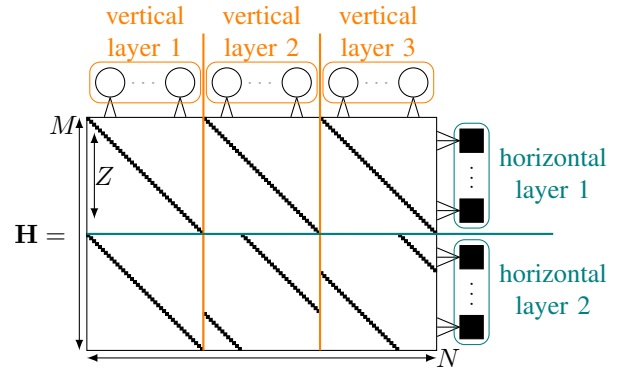

The horizontal scheme defines layers of check nodes that are fully updated, as shown in Fig.~\ref{fig:pcmqcldpcverticalhorizontal}~\cite{hocevar_reduced_2004}. Between the layer updates, all variable nodes  are partially updated, improving the reliability information for the next layers \emph{within} one iteration. In contrast, the vertical scheme defines layers of variable nodes that are fully updated, as shown in Fig.~\ref{fig:pcmqcldpcverticalhorizontal}~\cite{zhang_shuffled_2005}. Between the layer updates, all check nodes are partially updated.

This work investigates the two scheduling methods where compression operations are performed in each update under preservation of relevant information. The compression aims at reducing the message passing complexity and the memory footprint for caching messages between iterations. 

In mutual information maximizing decoding several implementations for the node updates exist\cite{mohr_coarsely_2021}. 
In this work we restrict ourselves to two-input operations that can be realized with standard components such as adders or comparators.
In~\cite{he_mutual_2019} a variable node with small single-input reconstruction tables, adders and non-uniform threshold quantization was shown to maximize the mutual information within a single node update. It was revealed in~\cite{mohr_uniform_2022} that restriction to uniform quantization allows significant complexity savings at nearly no performance loss. For the check node, solutions with non-uniform and uniform threshold quantization exist as well\cite{he_mutual_2019,mohr_uniform_2022}. Another check node implementation with slightly reduced performance but further complexity savings is the minimum approximation update~\cite{meidlinger_quantized_2015,mohr_uniform_2022}. The performance loss can be reduced when performing a check node aware variable node design as proposed in~\cite{mohr_variable_2022}.

The paper is organized as follows. Section \ref{sec:horizontalverticalarchitectures} describes the design of mutual information maximizing horizontal and vertical layered decoding. In section \ref{sec:complexity_analysis} the complexity for routing network and node update implementations are discussed. Section \ref{sec:performance} evaluates the error rate performance.

\section{Design of Quantized Layered Decoding}\label{sec:horizontalverticalarchitectures}
An LDPC encoder maps the information bits $\myvec{u}{\in} \{0,1\}^{K}$ to code bits $\myvec{b}{\in}\{0,1\}^{N}$ satisfying  $\underline{\mymat{H}}\myvec{b}{=}\myvec{0}$ where the parity check matrix $\underline{\mymat{H}}{\in} \{0,1\}^{M\times N}$ defines $M$ parity checks. 
Throughout the paper we assume a symmetrically quantized additive-white Gaussian noise (AWGN) channel with binary phase shift keying (BPSK) modulation. The quantization is designed to maximize the mutual information between the code bits $\myvec{b}$ and the $w_{ch}$-bit channel messages $\myvec{t}^{ch}{\in}\mathcal{T}_{w_{ch}}^{N}$~\cite{lewandowsky_information-optimum_2018}. All quantized messages use a symmetric sign-magnitude alphabet $\mathcal{T}_{i}{=}\{-2^{i-1},\ldots,-1,+1,\ldots,+2^{i-1}\}$, whose elements are sorted by the underlying log-likelihood ratio (LLR) $L(b|t){=}\log p(b{=}0|t)/p(b{=}1|t)$.
In the decoder we perform iterative message passing between variable and check nodes over a routing network.
To avoid routing congestion many applications make use of quasi-cyclic (QC) LDPC codes with a structured parity check matrix shown in Fig.~\ref{fig:pcmqcldpcverticalhorizontal}. The parity check matrix is fully defined by its base matrix $\mymat{H}{\in} \{-1,\ldots,Z-1\}^{N/Z\times M/Z}$ with lifting size $Z$: The lifting procedure replaces each element $h_{ij}$ with a cyclically shifted identity matrix $\mymat{I}(h_{ij}){\in} \{0,1\}^{Z\times Z}$ if $h_{ij}{\ge} 0$ and by a zero matrix $\mymat{0}{\in} \{0\}^{Z\times Z}$ if $h_{ij}{=}{-1}$\cite{mohr_coarsely_2021}. In this paper we restrict ourselves to regular LDPC codes with $h_{ij}\ge 0$ where $N/Z$ and $M/Z$ equal the check and variable node degrees $d_c$ and $d_v$.
The structure can be exploited to define horizontal or vertical layers, as highlighted in Fig. \ref{fig:pcmqcldpcverticalhorizontal} where $d_v{=}2$ and $d_c{=}3$. Next, the design of horizontal or vertical decoders is described, where we use discrete density evolution to track probability distributions of messages\cite{lewandowsky_information-optimum_2018}.
\subsection{Horizontal Layered Decoding}
One iteration in horizontal (or row-) layered decoding is characterized by $d_v$ layer updates. For each layer $l\in \mathcal{L}{=}\{0,\ldots,d_v{-}1\}$ we perform $Z$ full check node and $N$ partial variable node updates. Fig. \ref{fig:unrolledhorizontalcomp1} depicts a single unrolled iteration of the decoding procedure. 
\subsubsection{Full Check Node Updates}
We denote the set of variable nodes adjacent to check node $z{\in}\{0,\ldots,Z{-}1\}$ in layer $l$ as\begin{align*}
	\mathcal{V}_{l,z}{=}\left\{iZ{+}(z{+}h_{il}\Mod{Z})\ {:}\ i{\in} \{0,...,d_c{-}1\}\right\}.
\end{align*}
Each check node obtains extrinsic information for the variables through the parity check equation $b_{n}{=}\bigoplus_{n'{\in}\mathcal{V}_{l,z}\setminus\{n\}} b_{n'}$. For each connected variable node $n{\in}\mathcal{V}_{l,z}$, the mutual information maximizing check node update yields a $w$-bit output message $t^c_{l,n}{=}Q^c(y^c_{l,n}){\in} \mathcal{T}_{w}$ with
\begin{align}
y^c_{l,n}{=}\prod_{n'{\in}\mathcal{V}_{l,z}\setminus \{n\}}
\operatorname{sgn}(t^v_{l,n'})
\sum_{n'{\in}\mathcal{V}_{l,z}\setminus \{n\}} |\phi_c(t^v_{l,n'})|.
\label{equ:horizontal_comp}
\end{align}
In (\ref{equ:horizontal_comp}), $|\phi_c(t^v_{l})|{=}\log \tanh L(b_l|t^v_{l})$ is a reconstruction function, implemented by a small lookup table\cite{he_mutual_2019, mohr_uniform_2022}. 
The threshold quantization $Q^c$ poses as an information bottleneck (IB) setup where $y^c_{l,n}$, $b_n$ and $t^c_{l,n}$ are considered as the realizations of the observed, relevant and compressed random variables, $\myvar{Y}^c_{l}$, $\myvar{B}$ and $\myvar{T}^c_{l}$, respectively, with the objective $\max_{Q^c}I(\myvar{B}; \myvar{T}^c_{l})$~\cite{lewandowsky_information-optimum_2018}. Significant complexity can be saved through restriction to uniform quantization as proposed in~\cite{mohr_uniform_2022}. Alternatively, the minimum approximation update~\cite{mohr_coarsely_2021} yields
\begin{align}
t^c_{l,n}{=}\prod_{n'{\in}\mathcal{V}_{l,z}\setminus \{n\}}
\operatorname{sgn}(t^v_{l,n'})
\min_{n'{\in}\mathcal{V}_{l,z}\setminus \{n\}} |t^v_{l,n'}|.
\end{align}
\subsubsection{Partial Variable Node Update}
For $n{\in}\{0,\ldots,N{-}1\}$ the mutual information maximizing partial variable node update in layer $l$ yields a $w$-bit output message $t^v_{l{+}1,n} {=} Q^v(y^v_{l{+}1,n}){\in} \mathcal{T}_{w}$ with iterative or recursive computation of
\begin{align}
\begin{split}
y^v_{l{+}1,n}&{=}\phi_v(t^{ch}_n) + \sum_{l'\in \mathcal{L}\setminus\{l+1\}} \phi_v(t^c_{l',n})+L(b_n) \\ 
&=y^v_{l,n} + \phi_v(t^c_{l,n})-\phi_v(t^c_{l+1,n}).
\end{split}
\label{equ:partial_vnu_horizontal}
\end{align}
In (\ref{equ:partial_vnu_horizontal}), $\phi_v(t){=}L(t|b_n){=}\log p(t|b{=}0)/p(t|b{=}1)$ is a reconstruction function that can be implemented by small lookup tables and $L(b_n)=\log p(b_n{=}0/b_n{=}1)$ is the a priori LLR.
Note, that the index (de)increment $l{\pm1}$ is modulo $|\mathcal{L}|$. For the recursive update, we initialize $y^v_{0,n}{=}\phi_v(t^{ch}_n)$ in the first decoder iteration. Again, the non-uniform threshold quantization $Q^v$ poses an IB setup. A layer-specific design with low-complexity uniform threshold quantization is restricted to the iterative computation in (\ref{equ:partial_vnu_horizontal}) which allows rescaling of $y^v$~\cite{mohr_uniform_2022}. As shown in \cite{mohr_variable_2022} for the flooding schedule, a check node aware design of $Q^v$ may improve the performance also for the horizontal schedule.

\subsection{Vertical Layered Decoding}
One iteration in vertical (or column-) layered decoding is characterized by $d_c$ layer updates. For each layer $l\in \mathcal{L}{=}\{0,\ldots,d_c{-}1\}$ we perform $M$ partial check node and $Z$ full variable node updates. Fig. \ref{fig:unrolledverticalcomp} depicts a single unrolled iteration of the decoding procedure. In the first iteration we perform updates according to the flooding schedule with full check and variable node updates.
\subsubsection{Partial Check Node Updates}
For $m{\in}\{0,\ldots,M{-}1\}$ the mutual information maximizing partial check node update in layer $l$ yields a $w$-bit output message $t^c_{l,m} {=}Q^c(y^c_{l,m}){\in}\mathcal{T}_{w}$ with iterative or recursive computation of
\begin{align}
\begin{split}
y^c_{l,m}{=}&\prod_{l'{\in}\mathcal{L}\setminus \{l\}}
\operatorname{sgn}(t^v_{l',m})
\sum_{l'{\in}\mathcal{L}\setminus \{l\}} |\phi_c(t^v_{l',m})|\\
        =&\left(\operatorname{sgn}(y^c_{l{-}1,m}) \operatorname{sgn}(t^v_{l{-}1,m})\operatorname{sgn}(t^v_{l,m})\right)\cdot\\&\left(|y^c_{l{-}1,m}| {+} \phi_c(|t^v_{l{-}1,m}|){-}\phi_c(|t^v_{l,m}|)\right)\hspace{-0.1cm}.
\end{split}
\label{equ:partial_cnu_vertical}
\end{align}
Alternatively, the minimum approximation update yields
\begin{align}
t^c_{l,m}{=}\prod_{l'{\in}\mathcal{L}\setminus \{l\}}
\operatorname{sgn}(t^v_{l',m})
\min_{l'{\in}\mathcal{L}\setminus \{l\}} |t^v_{l',m}|.
\label{equ:vertical_min_approx}
\end{align}
Equation (\ref{equ:vertical_min_approx}) can be implemented with good accuracy using the  three-minimum approximation proposed in\cite{wang_reduced-complexity_2011}.
\subsubsection{Full Variable Node Update}
We denote the set of check nodes adjacent to variable node $z{\in}\{0,\ldots,Z{-}1\}$ in 
layer $l$ as
\begin{align*}
\mathcal{C}_{l,z}{=}\left\{iZ{+}(z{-}h_{il} \Mod{Z}) \ {:} \ i{\in} \{0,\ldots,d_v{-}1\}\right\}.
\end{align*}
For each connected check node $m{\in}\mathcal{C}_{l,z}$, the mutual information maximizing variable node update yields a $w$-bit output message $t^v_{l,m} {=} Q^v(y^v_{l{+}1,m}){\in}\mathcal{T}_{w}$ with
\begin{align}
\begin{split}
y^v_{l,m}&{=}\phi_{v}(t^{ch}_{lZ+z}) + \sum_{m'\in\mathcal{C}_{l,z}\setminus\{m\}} \phi_v(t^c_{l,m'})+L(b_{lZ+z}).
\end{split}
\end{align}

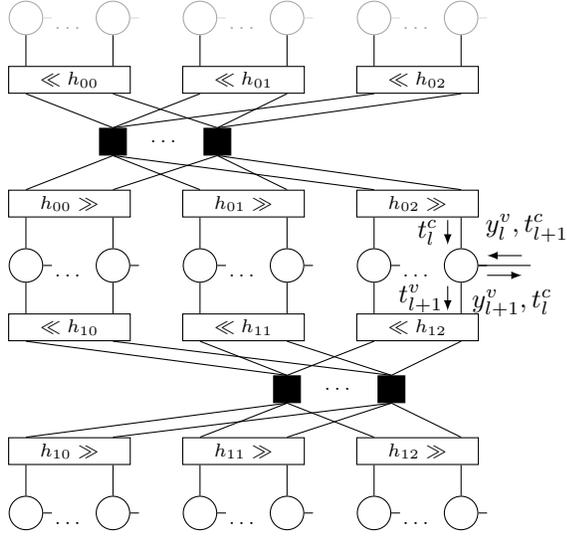
\begin{figure}[t]
	\centering
	% This file was created by tikzplotlib v0.9.8.
%    \resizebox{\figurewidth}{\figureheight}{
%	\tikzset{every picture/.style={line width=0.6pt}}
	\def\unit{1.78cm}  % here you set units used in image
	\begin{tikzpicture} [scale=1.0, x=1.3*\unit,y=1.85*\unit, node distance=1.5*\unit and 1.5*\unit,
		wire/.style={black},
		mygray/.style={color=gray!70, text=gray!70}]   
		\pgfmathsetmacro\nrows{2};		
		\pgfmathsetmacro\ncols{3};
		% create initial variable nodes
		\foreach \c in {1,...,\ncols}{
			\pgfmathsetmacro\r{0};		
			\pgfmathsetmacro\ycoord{-\r-0.5};		
			\node(vn1\r\c) at (\c,\ycoord)[draw, circle, minimum width=0.25*\unit, mygray]{};
			\node(dots\r\c) at (\c+0.25,\ycoord-0.04)[mygray]{\scriptsize$\dots$};
			\node(vnZ\r\c) at (\c+0.5,\ycoord)[draw, circle, minimum width=0.25*\unit, mygray]{};
			\draw[wire,gray!30](vn1\r\c.east)--++(0.05,0.0);
			\draw[wire,gray!30](vnZ\r\c.east)--++(0.05,0.0);
		}
		% create variable nodes off current iteration
		\foreach \r in {1,...,\nrows}{ % 0 is previous iteration		
			\foreach \c in {1,...,\ncols}{	
				\pgfmathsetmacro\ycoord{-\r-0.5};		
				\node(vn1\r\c) at (\c,\ycoord)[draw, circle, minimum width=0.25*\unit]{};
				\node(dots\r\c) at (\c+0.25,\ycoord-0.04)[]{\scriptsize$\dots$};
				\node(vnZ\r\c) at (\c+0.5,\ycoord)[draw, circle, minimum width=0.25*\unit]{};
				\draw[wire](vn1\r\c.east)--++(0.05,0.0);
				\draw[wire](vnZ\r\c.east)--++(0.05,0.0);
			}
		}
		% create check nodes and barrel shifters
		\foreach \r in {1,...,\nrows}{
			\node(cn1\r) at (0.5+\r,-\r)[draw, rectangle, fill=black,minimum width=0.2*\unit, minimum height=0.2*\unit]{};
			\node(cndots\r) at (0.5+\r+0.3,-\r)[]{\scriptsize$\dots$};
			\node(cnZ\r) at (0.5+\r+0.6,-\r)[draw, rectangle, fill=black,minimum width=0.2*\unit, minimum height=0.2*\unit]{};
			
			\foreach \c in {1,...,\ncols}{
				\pgfmathtruncatemacro\rm{\r-1};				
				\coordinate(tmp) at (vn1\rm\c.west|-cn1\r.east);
				\coordinate(A) at ([yshift=0.1*\unit]$(vn1\rm\c.west)!0.5!(tmp)$);
				\coordinate(tmp) at (vnZ\rm\c.east|-cn1\r.east);
				\coordinate(B) at ([yshift=-0.1*\unit]$(vnZ\rm\c.east)!0.5!(tmp)$);
				\pgfmathtruncatemacro\labelr{Mod(\r-1,\nrows)};	
				\pgfmathtruncatemacro\labelc{Mod(\c-1,\ncols)};	
				\node at ($(A)!0.5!(B)$)[rectangle, inner sep=2]{\scriptsize$\ll h_{\labelr\labelc}$};
				\node(barrel1\r\c)[draw, rectangle, fit=(A)(B), inner sep=0]{};
				\draw[wire](cn1\r.north)--(vn1\r\c|-barrel1\r\c.south);
				\draw[wire](cnZ\r.north)--(vnZ\r\c|-barrel1\r\c.south);
				
				\draw[wire](vn1\rm\c.south)--(vn1\rm\c.south|-barrel1\r\c.north);
				\draw[wire](vnZ\rm\c.south)--(vnZ\rm\c.south|-barrel1\r\c.north);
				
				\coordinate(tmp) at (vn1\r\c.west|-cn1\r.east);
				\coordinate(A) at ([yshift=0.1*\unit]$(vn1\r\c.west)!0.5!(tmp)$);
				\coordinate(tmp) at (vnZ\rm\c.east|-cn1\r.east);
				\coordinate(B) at ([yshift=-0.1*\unit]$(vnZ\r\c.east)!0.5!(tmp)$);
				
				\node at ($(A)!0.5!(B)$)[rectangle, inner sep=2]{\scriptsize$h_{\labelr\labelc}\gg$};
				\node(barrel2\r\c)[draw, rectangle, fit=(A)(B), inner sep=0]{};
				\draw[wire](cn1\r.south)--(vn1\r\c|-barrel2\r\c.north);
				\draw[wire](cnZ\r.south)--(vnZ\r\c|-barrel2\r\c.north);
				
				\coordinate(barrel2out1\r\c) at (vn1\r\c.north|-barrel2\r\c.south);
				\draw[wire](barrel2out1\r\c)--(vn1\r\c.north);
				\coordinate(barrel2outZ\r\c) at (vnZ\r\c.north|-barrel2\r\c.south);
				\draw[wire](barrel2outZ\r\c)--(vnZ\r\c.north);
			}
	}
%	\path(cnZ1.south)--node[midway, sloped, above]{\begin{tikzpicture}
%			\draw[latex-](0.0,0.0)--(0.2,0.0);
%	\end{tikzpicture}}++(0.3,0.0);
	\draw[wire](vnZ1\ncols.east)--node[midway, sloped, below]{\begin{tikzpicture}
			\draw[-latex](0.0,0.0)--(0.2,0.0);
	\end{tikzpicture}}node[yshift=+0.5cm, xshift=0.3cm]{$y^v_{l},t^c_{l{+}1}$}++(0.3,0.0);
	\path(vnZ1\ncols.east)--node[midway, sloped, above]{\begin{tikzpicture}
		\draw[latex-](0.0,0.0)--(0.2,0.0);
\end{tikzpicture}}node[yshift=-0.5cm, xshift=0.1cm]{$y^v_{l{+}1},t^c_{l}$}++(0.3,0.0);
	\draw[latex-]([xshift=-0.1*\unit, yshift=0.025*\unit]vnZ1\ncols.north)--node[yshift=0.0cm, xshift=-0.15*\unit]{$t^c_{l}$}++(0.0,0.1);
	\draw[-latex]([xshift=-0.1*\unit, yshift=-0.025*\unit]vnZ1\ncols.south)--node[yshift=0.0cm, xshift=-0.2*\unit]{$t^v_{l+1}$}++(0.0,-0.1);
	\end{tikzpicture}
	%}
	\caption{Unrolled iteration with horizontal layers.}
	\vspace{-0.2cm}
	\label{fig:unrolledhorizontalcomp1}

\end{figure}
\begin{figure}[h]
	\begin{subfigure}[b]{0.6\linewidth}
		\centering
		% This file was created by tikzplotlib v0.9.8.
%    \resizebox{\figurewidth}{\figureheight}{
%	\tikzset{every picture/.style={line width=0.6pt}}
	\def\unit{1cm}  % here you set units used in image
	\begin{tikzpicture} [scale=1.0, x=1.3*\unit,y=1.6*\unit, node distance=1.5*\unit and 1.5*\unit,
		wire/.style={black}]   
		\pgfmathsetmacro\nrows{2};		
		\pgfmathsetmacro\ncols{3};
		\pgfmathsetmacro\Z{8};
		% create variable node, memory and barrel shifter
		\foreach \c in {1,...,\ncols}{		
%			\pgfmathsetmacro\ycoord{-\r-0.5};
			\foreach \z in {1,...,\Z}{
				\pgfmathsetmacro\delta{0.9/\Z};
				\node(vn\z\c) at (0,-\c+1-\delta*\z)[draw, circle, minimum width=1.3*\delta*\unit, inner sep=0, color=black]{};
				\node(mem\z\c) at (0.3,-\c+1-\delta*\z)[draw, rectangle, fill=gray!50,minimum width=1.3*\delta*\unit, minimum height=1.3*\delta*\unit,inner sep=0]{};
				\draw[wire](vn\z\c)--(mem\z\c);	
		}
			\pgfmathtruncatemacro\labelc{\c-1};
			\coordinate(A) at ([xshift=-0.2*\unit, yshift=0.1*\unit]vn1\c.west);
			\coordinate(B) at ([xshift=-0.45*\unit, yshift=-0.1*\unit]vn\Z\c.west);
			\node at ($(A)!0.5!(B)$)[rectangle, inner sep=2,rotate=90]{\tiny$\ll h_{l\labelc}\gg$};
			\node(barrel\c)[draw, rectangle, fit=(A)(B), inner sep=0]{};
			\foreach \z in {1,...,\Z}{
				\draw[wire](vn\z\c.west)--(vn\z\c.west-|barrel\c.east);
			}
		}
		% create check nodes
		\foreach \z in {1,...,\Z}{
			\pgfmathsetmacro\delta{2/\Z};
			\node(cn\z) at (-\z*\delta-0.5,0.2)[draw, rectangle, fill=black,minimum width=0.1*\unit, minimum height=0.1*\unit]{};
			\foreach \c in {1,...,\ncols}{
				\pgfmathsetmacro\gap{0.06*\unit};
				\draw[wire]([xshift=2*\gap-\c*\gap]cn\z.south)|-(barrel\c.west|-vn\z\c);
			}
		}
	\node(labelatmem) at ([yshift=0.35*\unit]mem11)[inner sep=0]{\scriptsize memory};
	\draw[-latex](labelatmem)--(mem11);
	\draw[latex-latex]([xshift=0.15*\unit]mem11.north)--node[midway, right, xshift=-0.1*\unit]{\scriptsize$Z$}([xshift=0.15*\unit]mem\Z1.south);
%	\node(labelatmem) at ([yshift=0.3*\unit](mem11))[]{Mem.};
	\end{tikzpicture}
	%}
		\caption{2-dimensional view}
		\label{fig:horizontal_architecture_2d}    
	\end{subfigure}
	\begin{subfigure}[b]{0.37\linewidth}
		\centering
		\includestandalone[mode=image]{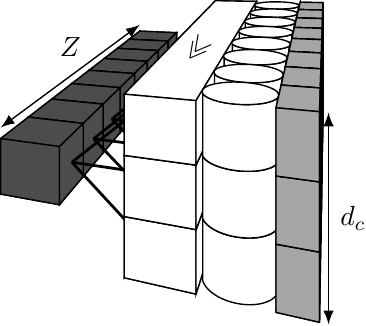}
		\caption{3-dimensional view}
		\label{fig:horizontalarchitecture3d}    
	\end{subfigure}
	\caption{Reorganized horizontal layer update ($Z{=}8$).}
	\label{fig:unrolled_horizontal}
\end{figure}
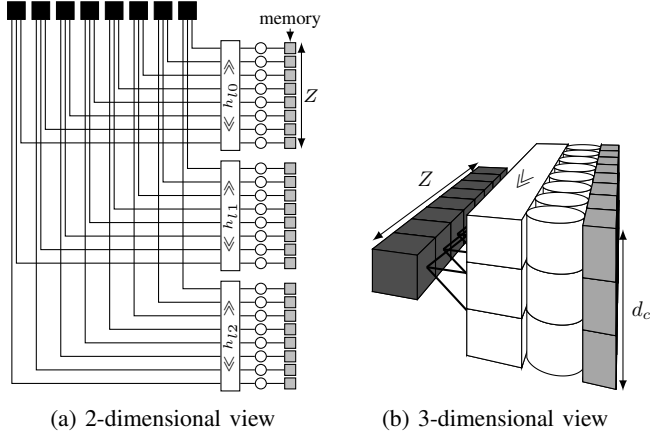
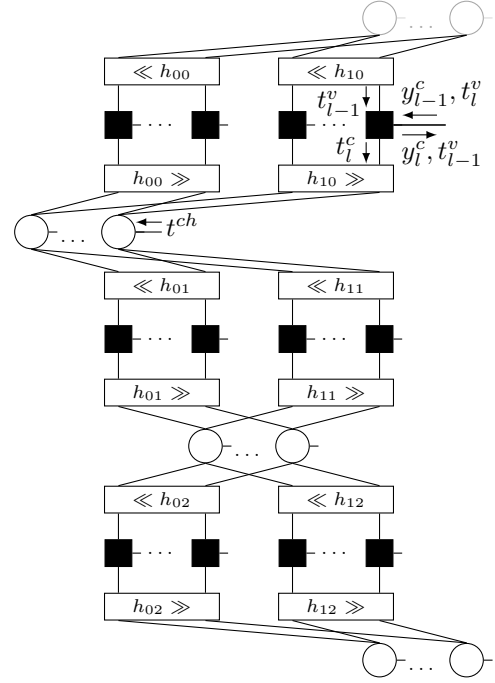
\begin{figure}[h]
	\centering
	% This file was created by tikzplotlib v0.9.8.
%    \resizebox{\figurewidth}{\figureheight}{
%	\tikzset{every picture/.style={line width=0.6pt}}
	\def\unit{1.78cm}  % here you set units used in image
	\begin{tikzpicture} [scale=1.0, x=1.3*\unit,y=1.6*\unit, node distance=1.5*\unit and 1.5*\unit,
		wire/.style={black},
		mygray/.style={color=gray!70, text=gray!70}]   
		\pgfmathsetmacro\nrows{2};		
		\pgfmathsetmacro\ncols{3};
		% create initial variable nodes
		\pgfmathsetmacro\c{0};		
		\pgfmathsetmacro\ycoord{-\c-0.5};		
		\node(vn1\c) at (\c+\ncols,\ycoord)[draw, circle, minimum width=0.25*\unit,mygray]{};
		\node(dots\c) at (\c+\ncols+0.25,\ycoord-0.04)[mygray]{\scriptsize$\dots$};
		\node(vnZ\c) at (\c+\ncols+0.5,\ycoord)[draw, circle, minimum width=0.25*\unit,mygray]{};
		\draw[wire,gray!30](vn1\c.east)--++(0.05,0.0);
		\draw[wire,gray!30](vnZ\c.east)--++(0.05,0.0);
		% create variable nodes off current iteration
		\foreach \c in {1,...,\ncols}{ % 0 is previous iteration		
			\pgfmathsetmacro\ycoord{-\c-0.5};		
			\node(vn1\c) at (\c,\ycoord)[draw, circle, minimum width=0.25*\unit]{};
			\node(dots\c) at (\c+0.25,\ycoord-0.04)[]{\scriptsize$\dots$};
			\node(vnZ\c) at (\c+0.5,\ycoord)[draw, circle, minimum width=0.25*\unit]{};
			\draw[wire](vn1\c.east)--++(0.05,0.0);
			\draw[wire](vnZ\c.east)--++(0.05,0.0);
		}
		% create check nodes and barrel shifters
		\foreach \c in {1,...,\ncols}{
			\foreach \r in {1,...,\nrows}{
				\node(cn1\c\r) at (0.5+\r,-\c)[draw, rectangle, fill=black,minimum width=0.2*\unit, minimum height=0.2*\unit]{};
				\node(cndots\r) at (0.5+\r+0.25,-\c)[]{\scriptsize$\dots$};
				\node(cnZ\c\r) at (0.5+\r+0.5,-\c)[draw, rectangle, fill=black,minimum width=0.2*\unit, minimum height=0.2*\unit]{};
				\draw[wire](cn1\c\r.east)--++(0.05,0.0);
				\draw[wire](cnZ\c\r.east)--++(0.05,0.0);
				
				\pgfmathtruncatemacro\labelr{Mod(\r-1,\nrows)};	
				\pgfmathtruncatemacro\labelc{Mod(\c-1,\ncols)};	
				% barrel shifter 1
				\pgfmathtruncatemacro\cm{\c-1};				
				\coordinate(tmp) at (cn1\c\r.west|-vn1\cm.west);
				\coordinate(A) at ([yshift=0.1*\unit]$(cn1\c\r.west)!0.5!(tmp)$);
				\coordinate(tmp) at (cnZ\c\r.east|-vnZ\cm.east);
				\coordinate(B) at ([yshift=-0.1*\unit]$(cnZ\c\r.east)!0.5!(tmp)$);
				\node at ($(A)!0.5!(B)$)[rectangle, inner sep=2]{\scriptsize$\ll h_{\labelr\labelc}$};
				\node(barrel1\c\r)[draw, rectangle, fit=(A)(B), inner sep=0]{};
				\draw[wire](cn1\c\r.north)--(cn1\c\r.north|-barrel1\c\r.south);
				\draw[wire](cnZ\c\r.north)--(cnZ\c\r.north|-barrel1\c\r.south);
				
				\draw[wire](vn1\cm.south)--(cn1\c\r.north|-barrel1\c\r.north);
				\draw[wire](vnZ\cm.south)--(cnZ\c\r.north|-barrel1\c\r.north);
				
				% barrel shifter 2
				\coordinate(tmp) at (cn1\c\r.west|-vn1\c.west);
				\coordinate(A) at ([yshift=0.1*\unit]$(cn1\c\r.west)!0.5!(tmp)$);
				\coordinate(tmp) at (cnZ\c\r.east|-vnZ\c.east);
				\coordinate(B) at ([yshift=-0.1*\unit]$(cnZ\c\r.east)!0.5!(tmp)$);
				\node at ($(A)!0.5!(B)$)[rectangle, inner sep=2]{\scriptsize$h_{\labelr\labelc}\gg$};
				\node(barrel2\c\r)[draw, rectangle, fit=(A)(B), inner sep=0]{};
				\draw[wire](cn1\c\r.south)--(cn1\c\r.south|-barrel2\c\r.north);
				\draw[wire](cnZ\c\r.south)--(cnZ\c\r.south|-barrel2\c\r.north);
				
				\draw[wire](vn1\c.north)--(cn1\c\r.south|-barrel2\c\r.south);
				\draw[wire](vnZ\c.north)--(cnZ\c\r.south|-barrel2\c\r.south);
				
			}
	}
\draw[wire](cnZ1\nrows.east)--node[midway, sloped, above]{\begin{tikzpicture}
			\draw[latex-](0.0,0.0)--(0.2,0.0);
	\end{tikzpicture}}node[yshift=+0.4cm, xshift=0.3cm]{$y^c_{l-1},t^v_{l}$}++(0.3,0.0);
\draw[wire](cnZ1\nrows.east)--node[midway, sloped, below]{\begin{tikzpicture}
		\draw[-latex](0.0,0.0)--(0.2,0.0);
\end{tikzpicture}}node[yshift=-0.4cm, xshift=0.3cm]{$y^c_{l},t^v_{l-1}$}++(0.3,0.0);
	\draw[latex-]([xshift=-0.1*\unit, yshift=0.025*\unit]cnZ1\nrows.north)--node[yshift=-0.1cm, xshift=-0.2*\unit]{$t^v_{l-1}$}++(0.0,0.1);
	\draw[-latex]([xshift=-0.1*\unit, yshift=-0.025*\unit]cnZ1\nrows.south)--node[yshift=0.05cm, xshift=-0.15*\unit]{$t^c_{l}$}++(0.0,-0.1);
	
	\draw[wire](vnZ1.east)--node[midway, sloped, above]{\begin{tikzpicture}
			\draw[latex-](0.0,0.0)--(0.15,0.0);
	\end{tikzpicture}}node[yshift=+0.1cm, xshift=0.45cm]{$t^{ch}$}++(0.15,0.0);
	\end{tikzpicture}
	%}
	\caption{Unrolled iteration with vertical layers.}
	\vspace{-0.3cm}
	\label{fig:unrolledverticalcomp}
\end{figure}
\begin{figure}[h]
    \begin{subfigure}[b]{0.6\linewidth}
    	\centering
    	% This file was created by tikzplotlib v0.9.8.
%    \resizebox{\figurewidth}{\figureheight}{
%	\tikzset{every picture/.style={line width=0.6pt}}
	\def\unit{1cm}  % here you set units used in image
	\begin{tikzpicture} [scale=1.0, x=1.3*\unit,y=1.6*\unit, node distance=1.5*\unit and 1.5*\unit,
		wire/.style={black}]   
		\pgfmathsetmacro\nrows{2};		
		\pgfmathsetmacro\ncols{3};
		\pgfmathsetmacro\Z{8};
		% create check nodes, memory and barrel shifter
		\foreach \r in {1,...,\nrows}{		
%			\pgfmathsetmacro\ycoord{-\r-0.5};
			\foreach \z in {1,...,\Z}{
				\pgfmathsetmacro\delta{0.9/\Z};
				\node(cn\z\r) at (0,-\r+1-\delta*\z)[draw, rectangle, minimum width=1.3*\delta*\unit,  minimum height=1.3*\delta*\unit, inner sep=0, fill=black, color=black]{};
				\node(mem\z\r) at (0.3,-\r+1-\delta*\z)[draw, rectangle, fill=gray!50,minimum width=1.3*\delta*\unit, minimum height=1.3*\delta*\unit,inner sep=0]{};
				\draw[wire](cn\z\r)--(mem\z\r);	
			}
			\pgfmathtruncatemacro\labelr{\r-1};
			\coordinate(A) at ([xshift=-0.2*\unit, yshift=0.1*\unit]cn1\r.west);
			\coordinate(B) at ([xshift=-0.45*\unit, yshift=-0.1*\unit]cn\Z\r.west);
			\node at ($(A)!0.5!(B)$)[rectangle, inner sep=2,rotate=90]{\tiny$\ll h_{\labelr l}\gg$};
			\node(barrel\r)[draw, rectangle, fit=(A)(B), inner sep=0]{};
			\foreach \z in {1,...,\Z}{
				\draw[wire](cn\z\r.west)--(cn\z\r.west-|barrel\r.east);
			}
		}
		% create variable nodes
		\foreach \z in {1,...,\Z}{
			\pgfmathsetmacro\delta{1.8/\Z};
			\node(vn\z) at (-\z*\delta-0.5,0.2)[draw, circle, minimum width=1.0*\delta*\unit,inner sep=0]{};
			\node(memch\z) at ([yshift=0.2*\unit]vn\z.north)[draw, rectangle, fill=gray!50,minimum width=0.75*\delta*\unit, minimum height=0.75*\delta*\unit,inner sep=0]{};
			\draw[wire](vn\z)--(memch\z);	
			\foreach \r in {1,...,\nrows}{
				\pgfmathsetmacro\gap{0.06*\unit};
				\draw[wire]([xshift=1.5*\gap-\r*\gap]vn\z.south)|-(barrel\r.west|-cn\z\r);
			}
		}
	\node(labelatmem) at ([yshift=0.35*\unit]mem11)[inner sep=0]{\scriptsize memory};
	\draw[-latex](labelatmem)--(mem11);
	\draw[latex-latex]([xshift=0.15*\unit]mem11.north)--node[midway, right, xshift=-0.1*\unit]{\scriptsize$Z$}([xshift=0.15*\unit]mem\Z1.south);
	%channel memory label
	\node(labelatmemch) at ($(memch\Z)!0.5!(memch1)$)[yshift=0.25*\unit, inner sep=0]{\scriptsize channel memory (read only)};
	\end{tikzpicture}
	%}
    	\caption{2-dimensional view}
    	\label{fig:verticalarchitecturecomp2}    
    \end{subfigure}
	\hspace{-0.7cm}
    \begin{subfigure}[b]{0.37\linewidth}
        \centering
		\includestandalone[mode=image]{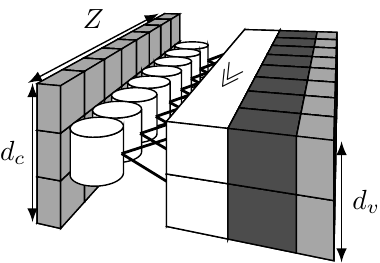}
        \caption{3-dimensional view}
        \label{fig:vertical_architecture_3d}    
    \end{subfigure}
    \caption{Reorganized vertical layer update ($Z{=}8$).}
    \label{fig:unrolled_vetical}
\end{figure}
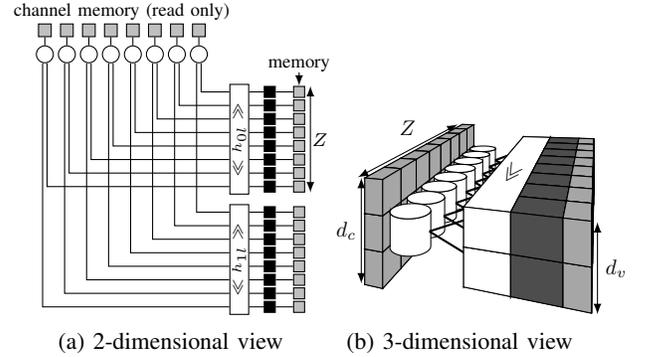
\section{Complexity Analysis}\label{sec:complexity_analysis}
\subsection{Routing Network Complexity}
For hardware implementations, the unrolled horizontal and vertical decoding graphs in Fig.~\ref{fig:unrolledhorizontalcomp1} and Fig.~\ref{fig:unrolledverticalcomp} can be reorganized to avoid routing congestion. In case of the horizontal schedule, Fig.~\ref{fig:horizontal_architecture_2d} places the variable and check nodes such that only parallel wires occur. The long parallel wires can be avoided by making use of the the third dimension. In Fig.~\ref{fig:horizontalarchitecture3d}, groups of $Z$ variable nodes are stacked on top of each other. In that configuration most of the routing complexity lies within the cyclic shifting units of size $Z$. Similarly, the vertical graph is reorganized in Fig.~\ref{fig:verticalarchitecturecomp2} and~\ref{fig:vertical_architecture_3d}. For the complexity analysis we assume the shifters to be implemented by reconfigurable barrel shifters that are realized with multiplexers. Alternatively, hardwired networks can be used. In Table \ref{table:barrel_shift_complexity} we depict the complexity of barrel shifters with values taken from \cite{boutillon_extended_2020}. Then, the complexity per shifted bit can be calculated by $3n_{\mylabel{mux}}/Z$ where we assume 3 logic gates per 2:1 multiplexer.  %
\begin{table}[!t]
	\centering
	\begin{tabular}{ c | c c c c c c }
		Z & 48 & 64 & 128 & 256 & 384 & 512 \\
		\hline
		$n_{\mylabel{mux}}$ & 336 & 448 & 1024 & 2304 & 3840 & 5120\\
		Gates per shifted bit& 21 & 21 & 24 & 27 & 30 & 30\\
	\end{tabular}
\caption{Barrel shifter complexity\cite{boutillon_extended_2020}. }
\vspace{-0.1cm}
\label{table:barrel_shift_complexity}
\end{table}
\vspace{-0.5cm}
\subsection{APP Message Passing}\label{sec:app_message_passing}
\begin{figure}[!t]
	\begin{subfigure}[b]{0.68\linewidth}
			\centering
			% This file was created by tikzplotlib v0.9.8.
	%    \resizebox{\figurewidth}{\figureheight}{
		%	\tikzset{every picture/.style={line width=0.6pt}}
		\def\unit{1.78cm}  % here you set units used in image
		\begin{tikzpicture} [scale=1.0, x=1.1*\unit,y=1.48*\unit, node distance=1.5*\unit and 1.5*\unit,
			wire/.style={black},
			mygray/.style={color=gray!70, text=gray!70},
			myred/.style={color=purple!40, text=purple!70},
			colorya/.style={color=purple!90, text=purple!90},
			coloryv/.style={color=teal!70, text=teal!90},
			hcircletop/.style={semicircle, shape border rotate=0,anchor=chord center,minimum width=0.25*\unit, inner sep=0},
			hcirclebot/.style={semicircle, shape border rotate=180,anchor=chord center,minimum width=0.25*\unit, inner sep=0}]   
			\pgfmathsetmacro\nrows{2};		
			\pgfmathsetmacro\ncols{3};
			\pgfmathsetmacro\ydelta{0.3};
			% create initial variable nodes
			\foreach \c in {1,...,\ncols}{
				\pgfmathsetmacro\r{0};		
				\pgfmathsetmacro\ycoord{-\r-0.0};		
				\node(tvn1\r\c) at (\c,\ycoord-\ydelta)[draw, hcircletop, mygray]{};
				\node(dots\r\c) at (\c+0.25,\ycoord-\ydelta)[color=gray!30]{\scriptsize$\dots$};
				\node(tvnZ\r\c) at (\c+0.5,\ycoord-\ydelta)[draw, hcircletop, mygray]{};
				\draw[wire,mygray](tvn1\r\c.east)--++(0.05,0.0);
				\draw[wire,mygray](tvnZ\r\c.east)--++(0.05,0.0);		
			}
			% create variable nodes off current iteration
			\foreach \r in {1,...,\nrows}{ % 0 is previous iteration		
				\pgfmathtruncatemacro\rm{\r-1};	
				\pgfmathtruncatemacro\rp{\r+1};	
				\pgfmathsetmacro\ycoord{-\r};
				\foreach \c in {1,...,\ncols}{	
					\node(bvn1\r\c) at (\c,\ycoord+\ydelta)[draw, hcirclebot, color=black]{};
					\node(dots\r\c) at (\c+0.25,\ycoord+\ydelta*1.2)[color=black]{\scriptsize$\dots$};
					\node(bvnZ\r\c) at (\c+0.5,\ycoord+\ydelta)[draw, hcirclebot, color=black]{};
					
					\node(tvn1\r\c) at (\c,\ycoord-\ydelta)[draw, hcircletop, color=black]{};
					\node(dots\r\c) at (\c+0.25,\ycoord-\ydelta*1.2)[color=black]{\scriptsize$\dots$};
					\node(tvnZ\r\c) at (\c+0.5,\ycoord-\ydelta)[draw, hcircletop, color=black]{};
					
					\coordinate(A) at ([yshift=0.1*\unit]$(tvn1\rm\c.west)!0.5!(bvn1\r\c.west)$);
					\coordinate(B) at ([yshift=-0.1*\unit]$(tvnZ\rm\c.east)!0.5!(bvnZ\r\c.east)$);
					
					\pgfmathtruncatemacro\labelrmm{Mod(\r-3,2)};	
					\pgfmathtruncatemacro\labelrm{Mod(\r-2,2)};	
					\node at ($(A)!0.5!(B)$)[rectangle, inner sep=2]{\tiny$\ll h_{\labelrmm\c}{-}h_{\labelrm\c}$};
					
%					\node at ($(A)!0.5!(B)$)[rectangle, inner sep=2]{\tiny$\ll h_{\rm\c}{-}h_{\r\c}$};
					\node(app barrel\r\c)[draw, rectangle, fit=(A)(B), inner sep=0]{};
%					\node(vn1\r\c) at (\c,\ycoord)[draw, circle, minimum width=0.25*\unit]{};
%					\node(dots\r\c) at (\c+0.25,\ycoord-0.04)[]{\scriptsize$\dots$};
%					\node(vnZ\r\c) at (\c+0.5,\ycoord)[draw, circle, minimum width=0.25*\unit]{};
%					\draw[wire](vn1\r\c.east)--++(0.05,0.0);
%					\draw[wire](vnZ\r\c.east)--++(0.05,0.0);
					\draw[wire,coloryv](bvn1\r\c.south)--(tvn1\r\c.north);	
					\draw[wire,coloryv](bvnZ\r\c.south)--(tvnZ\r\c.north);	
				}
				\node(cn1\r) at (0.5+\r,\ycoord)[draw, rectangle, fill=black,minimum width=0.2*\unit, minimum height=0.2*\unit]{};
				\node(cndots\r) at (0.5+\r+0.3,\ycoord)[]{\scriptsize$\dots$};
				\node(cnZ\r) at (0.5+\r+0.6,\ycoord)[draw, rectangle, fill=black,minimum width=0.2*\unit, minimum height=0.2*\unit]{};
				
			}
			% create next iteration vns with gray color
			\pgfmathtruncatemacro\r{\nrows+1};	
			\pgfmathtruncatemacro\rm{\r-1};	
			\pgfmathsetmacro\ycoord{-\r};
			\foreach \c in {1,...,\ncols}{	
				\node(bvn1\r\c) at (\c,\ycoord+\ydelta)[draw, hcirclebot, mygray]{};
				\node(dots\r\c) at (\c+0.25,\ycoord+\ydelta*1.2)[mygray]{\scriptsize$\dots$};
				\node(bvnZ\r\c) at (\c+0.5,\ycoord+\ydelta)[draw, hcirclebot, mygray]{};
				
				\coordinate(A) at ([yshift=0.1*\unit]$(tvn1\rm\c.west)!0.5!(bvn1\r\c.west)$);
				\coordinate(B) at ([yshift=-0.1*\unit]$(tvnZ\rm\c.east)!0.5!(bvnZ\r\c.east)$);
				\pgfmathtruncatemacro\labelrmm{Mod(\r-3,2)};	
				\pgfmathtruncatemacro\labelrm{Mod(\r-2,2)};	
				\node at ($(A)!0.5!(B)$)[rectangle, inner sep=2,mygray]{\tiny$\ll h_{\labelrmm\c}{-}h_{\labelrm\c}$};
				\node(app barrel\r\c)[draw, rectangle, fit=(A)(B), inner sep=0, mygray]{};
			}
			% wires
			\foreach \r in {1,...,\nrows}{
			\foreach \c in {1,...,\ncols}{
				\draw[wire,colorya](bvn1\r\c.north)--(bvn1\r\c.north|-app barrel\r\c.south);
				\draw[wire,colorya](bvnZ\r\c.north)--(bvnZ\r\c.north|-app barrel\r\c.south);
				\draw[wire](bvn1\r\c.south)--(cn1\r.north);
				\draw[wire](bvnZ\r\c.south)--(cnZ\r.north);
				\draw[wire](tvn1\r\c.north)--(cn1\r.south);
				\draw[wire](tvnZ\r\c.north)--(cnZ\r.south);
				\pgfmathtruncatemacro\rp{\r+1};	
				\draw[wire,colorya](tvn1\r\c.south)--(tvn1\r\c.south|-app barrel\rp\c.north);
				\draw[wire,colorya](tvnZ\r\c.south)--(tvnZ\r\c.south|-app barrel\rp\c.north);
				
				\draw[wire](tvn1\r\c.east)--++(0.05,0.0);
				\draw[wire](tvnZ\r\c.east)--++(0.05,0.0);
				\draw[wire](bvn1\r\c.east)--++(0.05,0.0);
				\draw[wire](bvnZ\r\c.east)--++(0.05,0.0);
			}
			}
			\foreach \c in {1,...,\ncols}{
				\pgfmathtruncatemacro\r{0};	
				\draw[wire, gray!30](tvn1\r\c.south)--(tvn1\r\c.north|-app barrel1\c.north);
				\draw[wire, gray!30](tvnZ\r\c.south)--(tvnZ\r\c.north|-app barrel1\c.north);
			}
			\foreach \c in {1,...,\ncols}{
				\pgfmathtruncatemacro\r{\nrows+1};	
				\draw[wire, gray!30](bvn1\r\c.north)--(bvn1\r\c.north|-app barrel\r\c.south);
				\draw[wire, gray!30](bvnZ\r\c.north)--(bvnZ\r\c.north|-app barrel\r\c.south);

				\draw[wire, gray!30](bvn1\r\c.east)--++(0.05,0.0);
				\draw[wire, gray!30](bvnZ\r\c.east)--++(0.05,0.0);
			}

			\draw[wire,coloryv](bvnZ1\ncols.south)--node[midway, sloped, below]{\begin{tikzpicture}
					\draw[-latex](0.0,0.0)--(0.2,0.0);
			\end{tikzpicture}}node[yshift=+0.0cm, xshift=-0.33cm]{$y^v_{l}$}(tvnZ1\ncols.north);
			
			\draw[wire](cnZ1.south)--node[midway, sloped, above]{\begin{tikzpicture}
					\draw[-latex](0.0,0.0)--(0.2,0.0);
			\end{tikzpicture}}node[yshift=+0.4cm, xshift=0.0cm]{$t^c_{l}$}(tvnZ1\ncols.north);
			
			\draw[wire](tvnZ1\ncols.east)--node[midway, sloped, above]{\begin{tikzpicture}
					\draw[-latex](0.0,0.0)--(0.2,0.0);
			\end{tikzpicture}}node[yshift=+0.4cm, xshift=0.0cm]{$t^c_{l}$}++(0.3,0.0);
			
			\draw[wire,colorya](tvnZ1\ncols.south)--node[yshift=-0.05cm,midway, sloped, above]{\begin{tikzpicture}
					\draw[-latex](0.0,0.0)--(0.15,0.0);
			\end{tikzpicture}}node[yshift=0.0cm, xshift=0.3cm,colorya]{$y^a_{l}$}(tvnZ1\ncols.south|-app barrel2\ncols.north);
			
			\draw[wire](bvnZ2\ncols.east)--node[midway, sloped, above]{\begin{tikzpicture}
				\draw[latex-](0.0,0.0)--(0.2,0.0);
			\end{tikzpicture}}node[yshift=+0.4cm, xshift=0.05cm]{$t^c_{l{+}1}$}++(0.3,0.0);
		
			\draw[wire](bvnZ2\ncols.south)--node[midway, sloped, below]{\begin{tikzpicture}
					\draw[latex-](0.0,0.0)--(0.2,0.0);
			\end{tikzpicture}}node[yshift=-0.45cm, xshift=0.1cm]{$t^v_{l{+}1}$}(cnZ2.north);
		
			\draw[wire,coloryv](bvnZ2\ncols.south)--node[midway, sloped, above]{\begin{tikzpicture}
					\draw[-latex](0.0,0.0)--(0.2,0.0);
			\end{tikzpicture}}node[yshift=+0.0cm, xshift=0.53cm]{$y^v_{l{+}1}$}(tvnZ2\ncols.north);
			
%			\path(vnZ1\ncols.east)--node[midway, sloped, above]{\begin{tikzpicture}
%					\draw[latex-](0.0,0.0)--(0.2,0.0);
%			\end{tikzpicture}}node[yshift=-0.5cm, xshift=0.1cm]{$y^v_{l{+}1},t^c_{l}$}++(0.3,0.0);
%			\draw[latex-]([xshift=-0.1*\unit, yshift=0.025*\unit]vnZ1\ncols.north)--node[yshift=0.0cm, xshift=-0.15*\unit]{$t^c_{l}$}++(0.0,0.1);
%			\draw[-latex]([xshift=-0.1*\unit, yshift=-0.025*\unit]vnZ1\ncols.south)--node[yshift=0.0cm, xshift=-0.2*\unit]{$t^v_{l+1}$}++(0.0,-0.1);
		\end{tikzpicture}
		%}
		\caption{Unrolled graph}
		\label{fig:horizontalarchitecturecompapp}   
	\end{subfigure}
	\begin{subfigure}[b]{0.31\linewidth}
		\centering
		\includestandalone[mode=image]{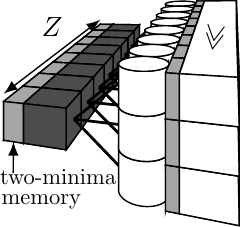}
		\caption{3-dimensional}
		\label{fig:horizontalarchitecturecompapp2}   
	\end{subfigure}
	\caption{APP message passing.}
	\label{fig:app_message_passing}
\end{figure}

The standard message passing in Fig.~\ref{fig:unrolledhorizontalcomp1} passes the compressed messages $t^c_l$ and $t^v_{l+1}$ through the shifting units $h_{13}\gg$ and $\ll h_{23}$. 
Alternatively, we can also transfer the a-posteriori probability (APP) message $y^a_{l}=y^v_{l}{+}\phi_v(t^c_l)$ with a single shift $(h_{13}{-}h_{23})\pmod Z$, which is an intermediate result of the partial variable node update (\ref{equ:partial_vnu_horizontal}). The APP message passing is more efficient if the bit width of $y^a_{l}$ is smaller than the combined bit width of $t^c_l$ and $t^v_{l{+}1}$. The horizontal APP unrolled decoding graph is depicted in Fig.~\ref{fig:horizontalarchitecturecompapp}. The variable and check node units are located closely without intermediate shifter as shown in Fig.~\ref{fig:horizontalarchitecturecompapp2}. Thus, using the minimum approximation allows to save memory by reconstructing $t^c_{l+1}$ from the first and second minimum (+index) of the previous iteration. 
Instead of storing $d_c (w-1)$ bits we only have $\lceil \log_2(d_c)\rceil {+} 2 (w-1)$ bits for the magnitudes. %
From (\ref{equ:partial_cnu_vertical}) we also can derive a modified APP message passing for the vertical schedule, but without further memory savings.

\subsection{Node Update Complexity}
Several options exist to implement the full and partial node update under a horizontal and vertical schedule. Table~\ref{table:node_update_complexity} gives an overview about the options for check node (CN) and variable node (VN). The update with non-uniform quantization was shown to achieve highest mutual information preservation~\cite{he_mutual_2019}. However, simulations in~\cite{mohr_coarsely_2021,mohr_uniform_2022} confirmed that approximations done with uniform quantization or the minimum approximation degrade the performance only by 0.01 or 0.05\,dB.

The last column counts the usage of reconstruction functions $\phi^c$ and $\phi^v$ which translate a $w$-bit message to a higher resolution $w'$-bit representation value. Note, that in uniform quantization~\cite{mohr_uniform_2022} the translation involves a scaling, such that the quantization can be achieved with a clipping and bit shifting operation. Thus, no comparisons or memory for storing the thresholds are required as in non-uniform quantization. One disadvantage of the uniform approach is that for every partial update all extrinsic inputs have to be rescaled and processed which is not very practical for high node degrees. The authors of~\cite{kang_memory_2022} observed that enforcing non-varying translation tables among consecutive layer updates causes only minor degradation. In a similar way the rescaling issue might be relaxed also for the uniform quantization.
In the following we focus on low-complexity configurations. 
\begin{table}[t]
	\centering
	\begin{tabular}{p{0.02cm} l l c c c}
		&\multicolumn{2}{c}{node type}  & additions & comparisons & translations \\
		\hline
		\parbox[t]{2mm}{\multirow{5}{0pt}{\rotatebox[origin=c]{90}{full}}}
		&\multirow{3}{0.2cm}{CN}&non-uniform\cite{he_mutual_2019} &$2d_c{-}2$&$d_c(w{-}1)$& $d_c$\\
		&&uniform\cite{mohr_uniform_2022} & $2d_c{-}2$& $0$ & $d_c{-}2$\\
		&&min\cite{meidlinger_quantized_2015}& $0$ & $d_c{+}\log_2(d_c){-}2$ & $0$\\\cline{3-6}
		&\multirow{2}{2pt}{VN}& non-uniform\cite{he_mutual_2019} & $2d_v{-}1$ & $d_v(w{-}1)$ & $d_v{+}1$ \\
		&&uniform\cite{mohr_uniform_2022}     & $2d_v{-}1$ & $0$ & $d_v{+}1$\\
		\hline
		\parbox[t]{2mm}{\multirow{5}{0pt}{\rotatebox[origin=c]{90}{partial}}}
		&\multirow{3}{0.2cm}{CN}& non-uniform&$2$&$w{-}1$& $1$\\
		&& uniform & $d_c{-}2$& $0$ & $d_c{-}2$\\
		&& 3-min\cite{wang_reduced-complexity_2011} & $0$ & $3$ & $0$ \\\cline{3-6}
		&\multirow{2}{2pt}{VN} &non-uniform & $2$ & $w{-}1$ & $1$ \\
		&&uniform& $d_v{-}1$ & $0$ & $d_v$
	\end{tabular}
	\caption{Node complexity depending on check node degree $d_c$, variable node $d_v$ and exchanged message resolution $w$.}
	\label{table:node_update_complexity}
\end{table}
\vspace{-0.05cm}
\subsection{Decoder Complexity}
Next, we aim to estimate the overall decoder resources including node computations, message transfers and memory demand for various \emph{practical} mutual information maximizing (MIM) decoders.
\begin{table}[t]
	\centering
	\begin{tabular}{ l c c c c c c c c c}
		bit width & 1 & 2 & 3 & 4 & 5 & 6 & 7 & 8 & 9 \\
		\hline
		gate count& 5  & 10 & 15& 20 & 25 & 30 & 35 & 40 & 45\\ 
	\end{tabular}
	\caption{Gate count per addition/comparison\cite{koren2018computer}.}
	\vspace{-0.3cm}
	\label{table:gate_counts_adder_units}
\end{table}
\begin{table}[t]
	\centering
	\begin{tabular}{ c|c c c c|c}
		\multirow{2}{*}{Decoder label}& CN ops. & VN ops. & Network & Total & Memory\\
		& [gates] & [gates] & [gates] & [gates] &[bits]\\
		\hline
		MIM-H 	  &\multirow{2}{*}{$11.2$}&$70$&$180$&$261$ &$2$ \\
		MIM-HA    &&$>70$&$240$&$>321$&$1.5$\\
		MIM-V   & $30$ &$58.3$ &$180$&$268$&$2.11$\\
		MIM-F 	  &$11.2$&$58.3$&$180$&$250$& 0\\
		\hline
		OMSQ-H		&\multirow{2}{*}{$18.5$}&$\multirow{2}{*}{60}$&$240$&$319$&$2.67$\\
		OMSQ-HA		&&&$180$&$259$&$1.61$\\
		OMSQ-V		&$45$ &$50$&$240$&$335$&$2.28$\\
		OMSQ-F		&$18.5$&$50$&$240$&$309$&0
	\end{tabular}
	\caption{Decoder complexity per edge in one iteration.}
	\label{table:estimated_decoder_complexity}
\end{table}
\subsubsection{Configurations}
For the horizontal layered decoder MIM-H we use a full check node update with minimum approximation and the partial variable node with uniform quantization. The decoder MIM-HA uses the APP message-passing schedule. For the vertical decoder MIM-V we select the partial check node update with 3-minimum approximation~\cite{wang_reduced-complexity_2011} and full variable node with uniform quantization~\cite{mohr_uniform_2022}. The flooding decoder MIM-F performs full updates for check and variable nodes. 
As a benchmark we consider the quantized offset-min-sum algorithm (OMSQ) decoder for flooding, horizontal-standard, horizontal-APP and vertical schedule\cite{zhang_shuffled_2005}. The check node uses a slightly more complex minimum approximation with offset operation.

The complexity for the addition and comparisons reported in Table \ref{table:gate_counts_adder_units} assumes a  $k$-bit ripple-carry adder with $5$ gates for each of the $k$ full-adders. This adder can be considered as a lower bound with minimum area but high delay\cite{koren2018computer}.

The MIM decoders use $w{=}3$-bit messages and $w'{=}6$\,bit-reconstructions. For the OMSQ decoder we consider $w{=}4$\,bits. In the complexity analysis and simulations we use a high rate code with $Z{=}512$, $d_v{=}3$ and $d_c{=}18$\cite{mohr_coarsely_2021}:
\begin{align}
	\begin{split}
		\scriptsize
		\mymat{H}{=}\left[\tiny\begin{smallmatrix}
			0& 0& 0& 0& 0& 0& 0& 0& 0& 0& 0& 0& 0& 0& 0& 0& 0& 0\\
			0& 205& 227& 29& 84& 427& 182& 116& 57& 332& 217& 308& 424& 363& 445& 439& 291& 368\\
			0& 327& 379& 458& 178& 105& 336& 162& 386& 212& 136& 109& 80& 198& 215& 289& 266& 204
		\end{smallmatrix}\right]
	\end{split}
\label{equ:code}
\end{align}
\subsubsection{Evaluation}
In Table \ref{table:estimated_decoder_complexity} the highest gate counts are related to the barrel-shifting routing network, which confirms that LDPC decoding is a data transfer dominated application, raising demand for low-resolutions.

The OMSQ decoders require $w{=}4$-bit messages since the representation levels cannot change across the iterations. But, the constant 4-bit levels also lead to $\lceil\log_2(d_v{\cdot}2^w)\rceil{=}6$-bit APP messages in the variable nodes. Therefore, under an APP message passing schedule, only 6\,bits for OMSQ-HA instead of $2{\cdot }4$\,bits for OMSQ-H must be transferred. 

The MIM decoders focus on reducing the resolution of messages under standard message passing. The reconstruction operation enables iteration-specific representation levels for the variable node update. Every reconstruction table consists of $2^{w{-}1}$ $w'{-}1$-bit values under a symmetric sign-magnitude format\cite{mohr_uniform_2022}. It can be observed in \cite{mohr_variable_2022} that only the high reliable magnitude level involve a non-linear translation. To keep the Table~\ref{table:estimated_decoder_complexity} less complicated, we have not included the iteration-dependent reconstruction complexity. 
The 6-bit reconstructions entail 7-bit adder units leading to 8-bit APP messages. 
Therefore, we have a larger routing network for MIM-HA compared to MIM-H. Further, the uniform quantization with bit shifting involves a rescaling of the APP message.
On the other hand, APP message passing reduces the memory demand from 2 to 1.5\,bits, as discussed in section \ref{sec:app_message_passing}. 

In the MIM-V decoder the uniform quantization is more efficient when fully updating the variable nodes. It avoids additional translations for internal rescaling of all extrinsic inputs that would be required in every partial update~\cite{mohr_uniform_2022}. Moreover, all check node messages can use the same reconstruction table. On the other hand, the 3-min check node updates are more complex.

The flooding decoders have the lowest node complexity by relying only on full node updates, however, they require twice the number of iterations compared to layered decoders\cite{mohr_coarsely_2021}.

\FloatBarrier
\section{Performance investigation}\label{sec:performance}
\begin{figure}[t]
	\centering
	\pgfplotsset{legend style={nodes={scale=0.7, transform shape}}}
    % This file was created by tikzplotlib v0.9.8.
\begin{tikzpicture}

\definecolor{color0}{rgb}{0.341176470588235,0.741176470588235,0.764705882352941}

\begin{axis}[
legend cell align={left},
legend style={
  fill opacity=0.8,
  draw opacity=1,
  text opacity=1,
  at={(0.01,0.01)},
  anchor=south west,
  draw=white!80!black
},
log basis y={10},
width=8cm,
height=7cm,
tick align=outside,
tick pos=left,
x grid style={white!69.0196078431373!black},
xlabel={\(\displaystyle E_b/N_0\) in dB},
xmajorgrids,
xmin=2.8, xmax=3.8,
xminorgrids,
xtick style={color=black},
y grid style={white!69.0196078431373!black},
ylabel={BER},
ymajorgrids,
ymin=1e-06, ymax=0.05,
yminorgrids,
ymode=log,
ytick style={color=black}
]
 \input{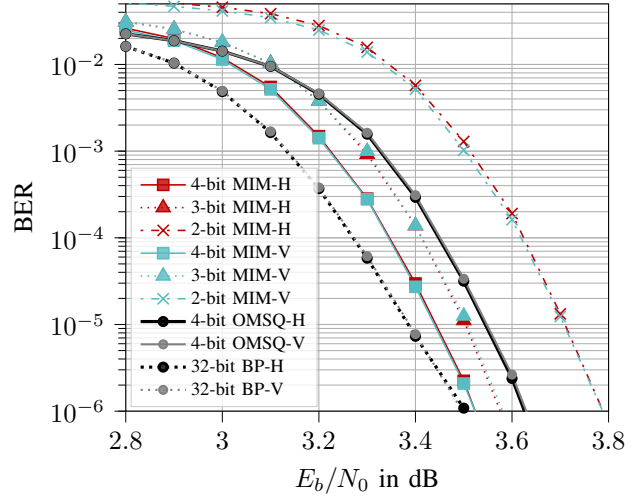}
\end{axis}

\end{tikzpicture}
	\vspace{-0.2cm}
	\caption{Bit error rate performance.}
	\vspace{-0.6cm}
	\label{fig:berrate562to4bit}
\end{figure}

\begin{figure}
	\centering
	    \pgfplotsset{
	% define the layers you need.
	% (Don't forget to add `main' somewhere in that list!!)
	layers/my layer set/.define layer set={
		background,
		main,
		foreground
	}{
		% you could state styles here which should be moved to
		% corresponding layers, but that is not necessary here.
		% That is why we don't state anything here
	},
	% activate the newly created layer set
	set layers=my layer set,
}
\pgfplotsset{legend style={nodes={scale=0.7, transform shape}}}
    % This file was created by tikzplotlib v0.9.8.
\begin{tikzpicture}

\definecolor{color0}{rgb}{0.341176470588235,0.741176470588235,0.764705882352941}

\begin{axis}[
tick align=outside,
tick pos=left,
height=5.5cm,
width=8cm,
x grid style={white!69.0196078431373!black},
xlabel={\(\displaystyle E_b/N_0\) in dB},
xmajorgrids,
xmin=3.3, xmax=4.3,
xminorgrids,
xtick style={color=black},
y grid style={white!69.0196078431373!black},
ymajorgrids,
ymin=2, ymax=6,%10.1,
yminorgrids,
ylabel={Avg. decoder iterations},
ytick style={color=black}
]
 \input{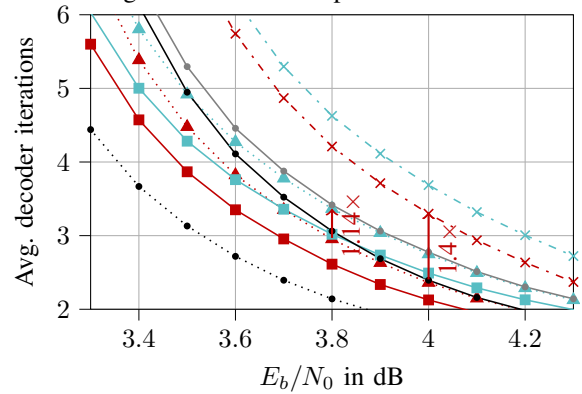}
 
%\node (A) at (3.8,4.6) {};
%\node (B) at (3.8,3.06) {};
%\draw[red!75.2941176470588!black,thick,->] (B) -- (A) node[midway,sloped,above,rotate=0] {1.3x};
\begin{pgfonlayer}{foreground}
\node (A) at (4,3.3) {};
\node (B) at (4,2.35) {};
\draw[red!75.2941176470588!black,thick,->,shorten >= -4pt, shorten <= -2pt] (B) -- (A) node[midway,sloped,below,rotate=0] {1.4$\times$};
\node (A2) at (3.8,2.95) {};
\node (B2) at (3.8,3.35) {};
\draw[->,thick,shorten >= -4pt, shorten <= -3pt,red!75.2941176470588!black] (A2) -- (B2) node[midway,sloped,below,rotate=0] {1.14$\times$};
\end{pgfonlayer}
\end{axis}

\end{tikzpicture}
	\vspace{-0.2cm}
	\caption{Average number of decoding iterations.}
	\vspace{-0.2cm}
	\label{fig:avgiterrate562to4bit1}
\end{figure}

In Fig.~\ref{fig:berrate562to4bit} we evaluate the bit error rate performance for the high-rate $R{=}5/6$ QC LDPC code (\ref{equ:code}) with a maximum of 10 iterations including results for high-resolution belief propagation (BP). 
Compared to 4-bit OMSQ-H, we observe gains of 0.1\,dB in case of 4-bit, 0.04\,dB for 3-bit and a degradation of 0.16\,dB for 2-bit MIM decoding. The horizontal and vertical decoder lead to very similar performance. 
In Fig.~\ref{fig:avgiterrate562to4bit1}, the vertical schedule requires 14\% more average decoding iterations (under early termination with the APP hard decision) compared to the 4-bit horizontal MIM decoder. Reducing the resolution from $w{=3}$ to $2$\,bits increases the average iteration count by 40\% at $E_b/N_0{=}4.0$\,dB, as highlighted. However, this is compensated with a smaller routing network which involves only 120 instead of 180 gates (Table~\ref{table:estimated_decoder_complexity}).

\section{Conclusions}
We compared horizontal and vertical layered decoding with mutual information maximizing node updates for regular quasi-cyclic LDPC codes. 
A complexity analysis revealed that barrel shifting constitutes a major part of the decoder. Decreasing bit width reduces shifting complexity but increases the average iteration count. Our results suggest less complexity for horizontal scheduling since we observed fewer average iterations. 
However, the schedule selection may depend on other important characteristics, like achievable clock frequency, required chip area or energy consumption. The evaluation of those metrics demands hardware implementations.

\bibliographystyle{MyIEEEtran}
\bibliography{literature}

\end{document}